# Kinetics of aqueous media reactions via ab initio enhanced molecular dynamics: the case of urea decomposition

Daniela Polino[a],[b] and Michele Parrinello*,[a],[b],[c]

[a]*Department of Chemistry and Applied Biosciences, ETH Zurich, c/o USI Campus Via Giuseppe Buffi 13, CH-6900 Lugano Switzerland*

[b]*Facoltà di Informatica, Istituto di Scienze Computazionali, Università della Svizzera Italiana, Via Giuseppe Buffi 13, CH-6900 Lugano Switzerland*

[c]*Istituto Italiano di Tecnologia, Via Morego 30, 16163 Genova, Italy*

E-mail: daniela.polino@phys.chem.ethz.ch, parrinello@phys.chem.ethz.ch

## Abstract

Aqueous solutions provide a medium for many important reactions in chemical synthesis, industrial processes, environmental chemistry, and biological functions. It is an accepted fact that aqueous solvents can be direct participants to the reaction process and not act only as simple passive dielectrics. Assisting water molecules and proton wires are thus essential for the efficiency of many reactions. Here we study the decomposition of urea into ammonia and isocyanic acid by means of enhanced ab initio molecular dynamics simulations. We highlight the role of the solvent molecules and their interactions with the reactants providing a proper description of the reaction mechanism and how the water hydrogen-bond network affects the reaction dynamics. Reaction free energy and rates have been calculated taking into account this important effect.

---

\* To whom correspondence should be addressed

# Introduction

The theoretical study of reaction kinetics in liquids is particularly challenging since solute-solvent interactions at various stages of the reactive pathway can influence the dynamic evolution of the reaction. Several methodologies with different degrees of approximation have been developed to address this issue. Among them, the most popular technique is based on high-level quantum mechanics (QM) calculations combined with continuum solvation models, such as the polarizable continuum model (PCM)[1]. These models are used to compute both equilibrium properties and reaction rates via transition state theory (TST)[2] or related approaches. When necessary, some reactant-solvent interactions can be considered by introducing a few explicit solvent molecules.

This methodology is computationally efficient and has proven to be successful in a large number of cases.[3] However, it fails when the solvent participates to the reaction, such as in proton-transfer type reactions, in which water may play a significant role. A case study for this class of reactions is urea (($NH_2$)$_2$CO) decomposition in aqueous solutions.

Due to electronic resonance effects, urea is stable in water. Nevertheless, understanding the mechanism and kinetics of its decomposition is important in a variety of applications that include agriculture[4], medical technologies[5] and energy.[6] Urea decomposition is also important in lean-burn automotive exhaust after treatment systems, since an urea-water solution is injected upstream of the deNOx catalyst to generate ammonia for the selective catalytic reduction (SCR) process.[7–9] Understanding the mechanism of urea decomposition is thus important in the engineering of such devices. Likewise, a comprehensive modelling of the kinetics of urea decomposition in aqueous solutions is necessary for the correct design of associated operation units in the wastewater treatment in chemical, food, and agricultural industries.[10,11]

In nature, urea is hydrolyzed by the urease enzyme, that is present in a variety of plants, selected fungi and bacteria species, yielding carbamic acid ($NH_2COOH$) and ammonia ($NH_3$).[12] However, Alexandrova and Jorgensen[13] have shown that the uncatalyzed decomposition of urea follows a different route, that leads to the formation of isocyanic acid (HNCO) and $NH_3$. This result is consistent with the experimental data of Kieke and coworkers.[14,15] In their work, they hydrolyzed urea at 473 K (200° C) and 275 bar in an FTIR spectroscopy flow reactor and detected the evolution of $CO_2$ and $NCO^-$ concentrations in time. The simplest kinetic scheme that fitted the experimental data, was based on the reactions:

R1) $(NH_2)_2CO \rightarrow NH_4^+ + NCO^-$

R2) $NH_4^+ + NCO^- + H_2O \rightarrow CO_2 + 2NH_3$

The correspondence found between the kinetic model proposed and the $CO_2$ and $NCO^-$ concentrations obtained from the hydro-thermolysis of urea confirms that the main decomposition route leads to HNCO.

Although the mechanism is not disputed, the kinetics of urea decomposition in water is still not fully understood. Theoretical studies[13,16–22] have shown, by means of high-level QM calculations coupled with implicit solvation models, that consistency between theory and experiment is approached only if one or more $H_2O$ molecules are added to the system. However, the corresponding rates calculated with standard TST deviate from experimentally measured ones by about one order of magnitude.

Water assistance is key to urea decomposition kinetics because it favors proton transfer between the two amino groups of urea, leading to the formation of HNCO and $NH_3$. Unfortunately, in high level electronic structure calculations, one can only afford the inclusion of a limited number of assisting water molecules. However, in solution proton transfer is enhanced by a more complex set of phenomena that involve the formation of hydrogen-bonds networks comprising many $H_2O$ molecules. This feature must be considered if one wants to describe correctly the catalytic role of the solvent and estimate its effective reaction rate.

Considering this, ab-initio molecular dynamics at the all-atom level represents the best tool for describing the reaction properties of these systems. Unfortunately for big systems the typical time scale achievable in ab-initio MD is only on the order of hundreds of picoseconds, not long enough to explore reactions, whose time scale can range from microseconds to hours. This inherent time scale limitation can be addressed by accelerating standard MD sampling by means of enhanced sampling methods, such as metadynamics.[23,24] A recent review of metadynamics can be found for instance in ref [25]. Within this scheme, a new tool, called infrequent metadynamics,[26] has been designed specifically to estimate the rate of rare events. In short, knowing the time of simulation necessary to go from one metastable state A to another metastable state B in a metadynamics run and the bias deposited to overcome the barrier separating the two, one can recover the corresponding physical transition time. In the rare event regime the observed transition times can be fitted to a Poisson distribution, from which the characteristic time can be extracted.[27] Therefore, combining ab-initio MD with infrequent metadynamics, one can estimate the characteristic time of a chemical reaction including in the calculation a proper description of the entropic contribution due to the concerted motion of the solvent molecules and the role that solute-solvent interaction has on the transition dynamics.

In this work we used this combined approach to determine the kinetics of urea decomposition in aqueous solutions. We thus provide a more detailed description of the decomposition mechanism, revealing how the network of $H_2O$ molecules affects the reaction dynamics. Additionally, we demonstrate that in liquid-phase reactions the reactive events cannot be defined by a single transition state geometry, but the fluctuations of the solvent, that are an integral part of the reaction coordinate, generate an ensemble of transition geometries. Infrequent metadynamics, which does not need any information about the transition state geometry, results as the most suitable approach to estimate the rate of such reactions.

## Methodology

Metadynamics[23] belongs to a class of methods in which the rare event sampling is enhanced by introducing a history dependent bias potential to a set of collective variables (CVs). In particular we used its well-tempered variant,[24] in which the height of the deposited Gaussian decreases exponentially as the bias is deposited:

$$V_n(s) = \sum_{k=1}^{n} G(s, s_k) e^{-\beta/(\gamma-1)V_{k-1}(s_k)} \tag{1}$$

Where $G(s, s_k) = W\, e^{-\frac{\|s-s_k\|^2}{2\sigma^2}}$ is a Gaussian centered in $s_k$, the CVs value at time $t_k$, $\beta = 1/k_B T$, $\gamma$ is the bias factor, and $\sigma$ is the width of the Gaussian. The choice of a limited number of relevant CVs is nontrivial, especially when the reaction of interest is characterized by a complex mechanism in which many degrees of freedom are involved. Thus, the design of CVs that fully include the important role of the solvent degrees of freedom is crucial for liquid-phase reactions.

Inspired by the work of Pietrucci and Saitta[28], which is based on the path-CVs concept introduced by Branduardi et al.[29], we defined two CVs, s and z, that are permutation invariant with regard to the title reaction. Briefly, we label as A and B the reactant (urea) and product (HNCO+$NH_3$) state, respectively, and design s and z so as to have the values s=-1 and z=1 in state A and s=1 and z=1 in state B:

$$s(\mathbf{R}) = \frac{-e^{-\lambda D[\mathbf{R},\mathbf{R}_A]} + e^{-\lambda D[\mathbf{R},\mathbf{R}_B]}}{e^{-\lambda D[\mathbf{R},\mathbf{R}_A]} + e^{-\lambda D[\mathbf{R},\mathbf{R}_B]}} \quad (2)$$

$$z(\mathbf{R}) = -\frac{1}{\lambda}\log\left(\frac{e^{-\lambda D[\mathbf{R},\mathbf{R}_A]} + e^{-\lambda D[\mathbf{R},\mathbf{R}_B]}}{2}\right) \quad (3)$$

with

$$D[\mathbf{R},\mathbf{R}_A] = (C_{CN} - C_{CN}^A)^2 + (\Delta C_{NH} - \Delta C_{NH}^A)^2 \quad (4)$$

$$D[\mathbf{R},\mathbf{R}_B] = (C_{CN} - C_{CN}^B)^2 + (\Delta C_{NH} - \Delta C_{NH}^B)^2 \quad (5)$$

$$\Delta C_{NH} = |C_{N1H} - C_{N2H}| \quad (6)$$

where $C_{CN}$ is the number of nitrogen atoms around the carbon, and $C_{N1H}$ and $C_{N2H}$ are the numbers of hydrogens around nitrogen N1 and N2 respectively. See supporting information (SI) for more details. The reference values for $C_{CN}$ and $\Delta C_{NH}$ are 2 and 0 in state A and 1 and 2 in state B (see Scheme 1). The parameter λ was chosen equal to 0.5.

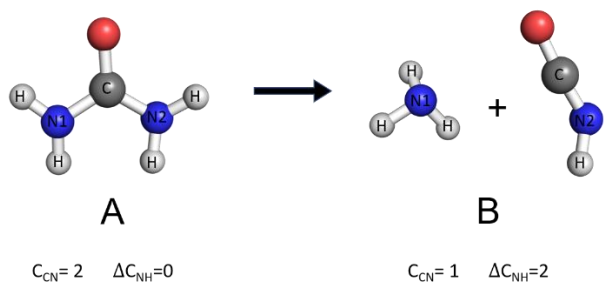

Scheme 1. Reference structure parameters for the definition of the path variables used in this work.

The CV s(R) measures how close or how distant configuration R is from the reactant or the product, while z(R) measures how distant one is from the linear path connecting A to B.

Ab initio molecular dynamics simulations were performed using Quickstep, which is part of the CP2K package[30] supplemented by the PLUMED 2.4 plugin.[31] In these calculations, the Born–Oppenheimer forces were used to propagate the nuclear dynamics. A convergence criterion of 5 × 10$^{-7}$ a.u. was used for the optimization of the wave function. The popular PBE exchange-correlation density functional[32] was employed to evolve the dynamics of a 10×10×10 Å box with 1 molecule of urea and 34 molecules of water with periodic boundary conditions. We used the m-DZVP (2s2p1d/2s1p) Gaussian basis set and a plane wave cutoff of 300 Ry for the density. Core electrons were treated using the Goedecker–Teter–Hutter (GTH) pseudopotentials.[33,34] The time step used was 1 fs. Production runs were carried out at 390 and 490 K in the NVT ensemble using the stochastic velocity-rescaling thermostat of Bussi et al.[35] In the former case we could not satisfactorily converge the FES but we were still able to evaluate the rates, with the method to be described below.

## Results and Discussion

A first explorative run was carried out at 490 K, employing a bias factor $\gamma = 40$ and deposition rate of 50 fs. The Gaussians adopted had an initial height of 1 kJ/mol and width of 0.1 and 0.2 for the s and z collective variables respectively. The system was first equilibrated at the chosen temperature with a MD simulation of 15 ps.

The free energy surface (FES) computed and the reaction mechanism extracted from the simulation are illustrated in Figure 1. As found by Alexandrova and Jorgenesen,[13] Urea (A) decomposition first passes through the formation of a metastable zwitterion intermediate (A1) which is 56.6 kJ/mol higher than state A. After the zwitterionic transient the C-$NH_3$ bond breaks forming ammonia and isocyanic acid, state B in Figure 1, which is 20.7 kJ/mol above urea. In water ammonia and isocyanic acid easily convert to the more stable $NCO^-$ and $NH_4^+$ ions, and as expected, also in our metadynamics run the system rapidly evolved from state B to state B1, which is about 9.7 kJ/mol higher in energy than urea.

Metadynamics, together with many other enhanced sampling approaches, focuses on the reconstruction of equilibrium properties. Therefore, information about the dynamics of the reactive event is lost. From a chemical point of view, we can say that metadynamics can provide good estimates of reaction free energies but not good enough values for the free energy barriers since the TS is sparsely sampled.[36] Infrequent metadynamics[26] uses suitably engineered well-tempered metadynamics runs to evaluate transition rates, i.e. free energy barriers. It is essential that during such runs no bias is deposited in the transition region such that the rapid well-to-well dynamics remains unaffected. To satisfy this request, infrequent metadynamics runs are characterized by very slow deposition strides and small initial Gaussian heights.

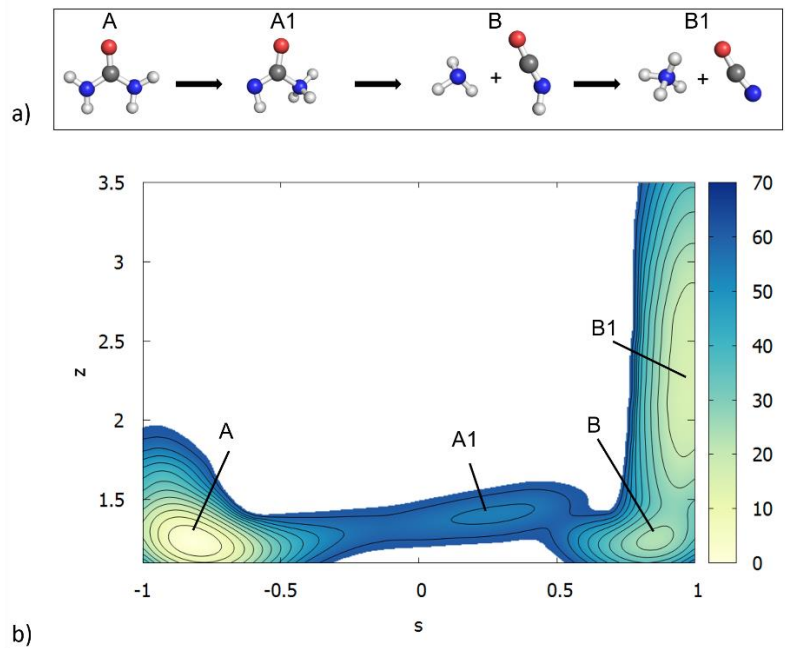

Figure 1. a) reaction mechanism b) Free energy surface (kJ/mol) of urea decomposition in water as a function of path collective variables s and z.

In our case, the energy barrier of urea decomposition in water is known to be in the range of 120-135 kJ/mol. As such, standard infrequent metadynamics would employ long simulation times to obtain accurate reaction times. Therefore, we drew inspiration from the recently developed variational flooding formalism[37]. In this approach one guarantees, using a variational procedure, that the free energy barriers remain bias-free. We combined this key element of variational flooding with infrequent metadynamics[26] as follows. First, we carried out a metadynamics run to fill the reactant basin (state A) up to a threshold preassigned free energy. The threshold was fixed at 90 kJ/mol since the first reactive event during the metadynamics run was detected after depositing 120 kJ/mol and experimental evidences set the activation barrier to urea decomposition in the range of 120-135 kJ/mol. The added bias was then used as an external potential and on this biased energy surface an infrequent metadynamics run was performed.

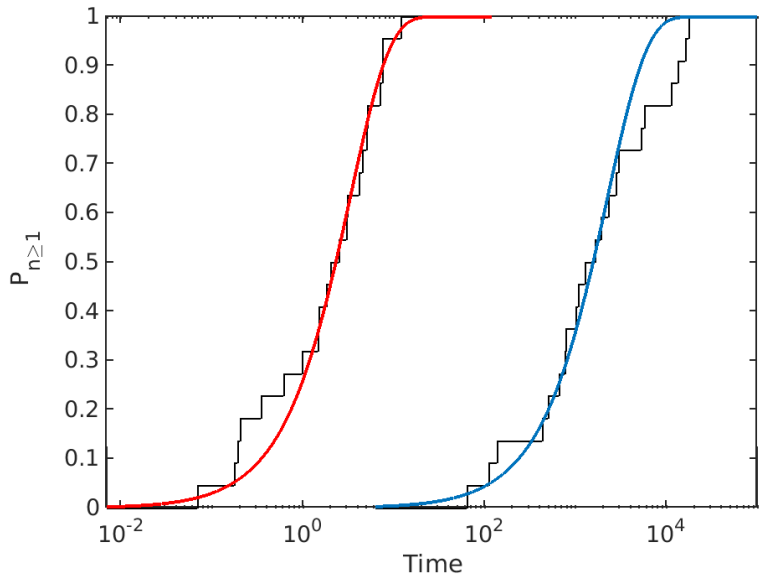

Figure 2. CDF and relative fit to a Poisson distribution function of the transition times for urea decomposition at 490 (red line) and 390 K (blue line).

In particular we evolved 30 independent trajectories, all initiated after thermalization in the A basin, adopting a large deposition stride and a small Gaussian height (1 ps and 0.4 kJ/mol respectively) as the infrequent metadynamics protocol requires. In order to satisfy the condition on which infrequent metadynamics is based, it is necessary to verify that even after the imposition of the bias one is still in the rare event regime. Under this condition, the distribution of escape times must be Poissonian and the statistical accuracy of this assumption verified.[27] This is shown in Figure 2. The p-test value found was 0.46 and 0.85, at 390 and 490 K respectively, assessing thus the reliability of the reconstructed dynamics. The decomposition rates at 390 and 490 K were found to be $4.46 \times 10^{-4}$ and $3.02 \times 10^{-1}$ s$^{-1}$.

We performed also a standard harmonic TST calculation[2] to have data to compare our urea decomposition rate to. The transition state geometry was determined considering the assistance of one water molecule and can be found in the supporting material. All the static electronic energy calculations were performed with the Gaussian 09 software package.[38] The activation energy obtained at the PBE/cc-pVDZ/IEFPCM level[32,39,40] is 83.1 kJ/mol and the reaction rate expressed in the Arrhenius form is $k=2.91 \times 10^{10} \exp(-83.1(\text{kJ/mol})/RT)$.

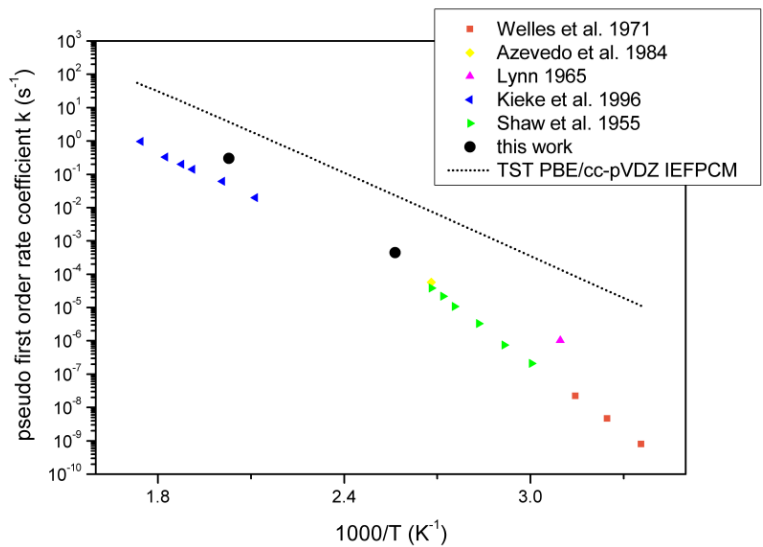

Figure 3. Pseudo first order rate constant of urea decomposition in aqueous media as a function of temperature calculated with standard TST (dashed black line), infrequent metadynamics (black circles) are compared with various experimental data (points).

The experimental pseudo first order rate constants[14,41–44] are drawn in Figure 3 as a function of temperature together with the standard TST estimation and our calculations. As shown, the rate obtained via standard TST is almost two orders of magnitude higher than the experimental one. The rate determined using infrequent metadynamics with explicit inclusion of the solvent dynamics at the PBE level is closer to the experiment in the temperature regime investigated (490-390 K). We hypothesize that the close agreement is due to error cancellations together with a more accurate description of the anharmonicity of the system.

More interestingly our calculations reveal that the urea decomposition mechanism is more complex than the model underpinning the TST calculations. In fact, the critical step in the reaction mechanism of urea decomposition is the formation of the zwitterion intermediate. The formation of the zwitterion ($^{-}$HNCONH$_3^{+}$) implies a proton transfer between urea's two nitrogen atoms. However, this event is not always instantaneous as represented by the single transition state found with standard static implicit solvent calculations. Actually, the system can react through an ensemble of transition states after a number of failed attempts. This leads to a large discrepancy between the TST estimation and that obtained with infrequent metadynamics.

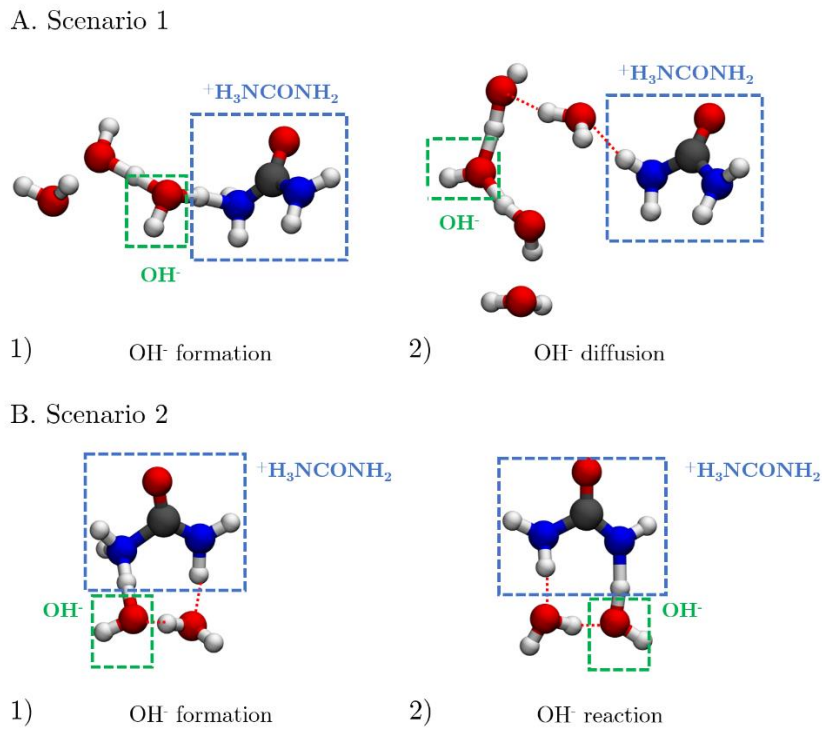

Figure 4. Different scenarios encountered in the formation of $^-$HNCONH$_3^+$. Formation and diffusion of OH$^-$ (A). Formation and rapid reaction of OH$^-$ with NH$_2$CONH$_3^+$ (B).

By harvesting a number of reactive trajectories we were able to identify two possible scenarios, as illustrated in Figure 4, where representative snapshots of these two scenarios are shown.

The first scenario involves an intermediate step as depicted in Figure 4a. The proton transfer passes first through the formation of the OH$^-$ + NH$_2$CONH$_3^+$ system, where the OH$^-$ diffuses away from the reactant. This diffusion happens via the formation of a proton wire that may involve up to 7 water molecules. After an average lifetime of 1 ps the OH$^-$ recombines with NH$_2$CONH$_3^+$ to form $^-$HNCONH$_3^+$ + H$_2$O. Interestingly, in this part of the simulation we observed water making several frustrated attempts to cede the proton. This reflects the diffusive nature of the transition region which is frequently encountered in liquid phase reactions and that TST static calculations fail to represent.

The second scenario is closer to the one represented by the static transition state as can be noticed in Figure 5b. In this case, in fact, the OH$^-$ does not diffuse away but it reacts almost instantaneously with the NH$_2$CONH$_3^+$ species thanks to the formation of a proton wire that connects directly the two nitrogen atoms. The life time of the OH$^-$ species in this situation is much shorter, 400 fs on average. This analysis exposes the complexity of these processes, in which the solvent is actively involved in the reaction and the kinetics cannot be correctly determined by one single transition state.

## Conclusions

The present methodology provides a powerful tool for the estimation of reaction times and, thus, reaction rates in condensed phase systems. As illustrated by the urea example the explicit consideration of the solute-solvent interactions in the estimation of proton transfer rates is essential for the correct description of the reaction dynamics. We conclude that the method is a promising tool for studying chemical reactions in liquid solutions and can be applied to several problems of industrial and biological interest.

## Acknowledgments

We thankfully acknowledge the financial support provided by CASALE SA, Via Giulio Pocobelli 6, 6900 Lugano, Switzerland and by the European Union Grant No. ERC-2014-AdG-670227/VARMET. Computational resources were provided by the Swiss National Supercomputing Centre (CSCS) under project ID s768, and by the ETH Zürich Euler cluster.

## Associated Content

Supporting Information: Details about the CVs used to perform our metadynamics calculations and convergence analysis of the metadynamics run used to recover the free energy surface illustrated in Figure 1. Details about the method adopted to perform TST calculations together with geometries, rotational constant and frequencies of the relevant structures.

**Table of Content Graphic**

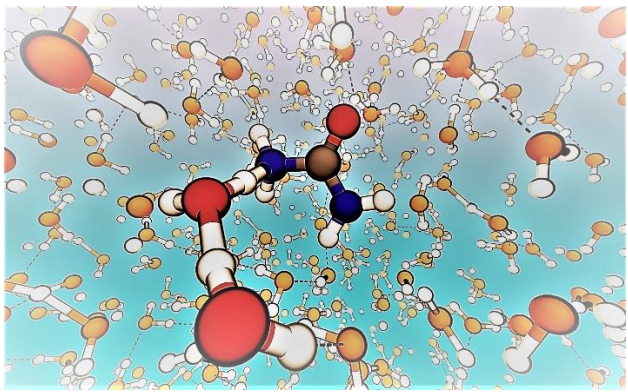

# Kinetics of aqueous media reactions via ab initio enhanced molecular dynamics: the case of urea decomposition

# Supporting Information

Daniela Polino and Michele Parrinello[*]

*Department of Chemistry and Applied Biosciences, ETH Zurich*
*Facoltà di Informatica, Istituto di Scienze Computazionali, Università della Svizzera Italiana,*
*Via G. Buffi 13, 6900 Lugano Switzerland*

**Table of contents**

---

[*] To whom correspondence should be addressed

# 1 Methodology

## 1.1 Metadynamics

To perform our metadynamics calculations we used the following collective variables (CVs), s and z, that are permutation invariant with regard to the title reaction. Briefly, we label as A and B the reactant (urea) and product (HNCO+NH$_3$) state, respectively, and design s and z so as to have the values s=-1 and z=1 in state A and s=1 and z=1 in state B:

$$s(\mathbf{R}) = \frac{-e^{-\lambda D[\mathbf{R},\mathbf{R}_A]} + e^{-\lambda D[\mathbf{R},\mathbf{R}_B]}}{e^{-\lambda D[\mathbf{R},\mathbf{R}_A]} + e^{-\lambda D[\mathbf{R},\mathbf{R}_B]}} \tag{S1}$$

$$z(\mathbf{R}) = -\frac{1}{\lambda}\log\left(\frac{e^{-\lambda D[\mathbf{R},\mathbf{R}_A]} + e^{-\lambda D[\mathbf{R},\mathbf{R}_B]}}{2}\right) \tag{S2}$$

with

$$D[\mathbf{R},\mathbf{R}_A] = \left(C_{CN} - C_{CN}^A\right)^2 + \left(\Delta C_{NH} - \Delta C_{NH}^A\right)^2 \tag{S3}$$

$$D[\mathbf{R},\mathbf{R}_B] = (C_{CN} - C_{CN}^B)^2 + (\Delta C_{NH} - \Delta C_{NH}^B)^2 \tag{S4}$$

$$\Delta C_{NH} = |C_{N1H} - C_{N2H}| \tag{S5}$$

The coordination numbers $C_{CN}$, $C_{N1H}$, and $C_{N2H}$ have been calculated as follows using a rational switching function to make the CV differentiable:

$$C_{CN} = \sum_{i \in C} \sum_{j \in N1,N2} \frac{1 - \left(\frac{r_{ij}}{r_0}\right)^n}{1 - \left(\frac{r_{ij}}{r_0}\right)^m} \tag{S6}$$

$$C_{N1H} = \sum_{i \in N1} \sum_{j \in H} \frac{1 - \left(\frac{r_{ij}}{r_0}\right)^n}{1 - \left(\frac{r_{ij}}{r_0}\right)^m} \tag{S7}$$

$$C_{N2H} = \sum_{i \in N2} \sum_{j \in H} \frac{1 - \left(\frac{r_{ij}}{r_0}\right)^n}{1 - \left(\frac{r_{ij}}{r_0}\right)^m} \tag{S8}$$

Here $r_{i,j}$ is the distance between atom $i$ and atom $j$. The parameters of the switching functions chosen are $r_0$=1.8 Å n=6 and m=12 for $C_{CN}$, and $r_0$=1.5 Å n=6 and m=12 for $C_{N1H}$ and $C_{N2H}$.

## 1.2 Static Calculations

To assess and compare our metadynamics simulations we performed static ab initio calculations. Geometries of the main reactant, products and transition state, were optimized using DFT at the B3LYP/6-31+G(d,p) level and adopting the Polarizable Continuum Model (PCM) using the integral equation formalism variant (IEFPCM) to account for solvent effects. The transition state was located using a synchronous transit guided saddle point search algorithm and characterized by a single imaginary frequency. All energies were then calculated at the PBE/cc-pVDZ level to compare rates using the same level of theory of the ab initio molecular dynamics simulation. The rate coefficient of urea decomposition was determined using standard transition state theory.

# 2 Results

## 2.1 Metadynamics

In Figure 1 is illustrated the behavior of the carbon-nitrogen coordination number ($C_{CN}$) as a function of simulation time, to show the ability of the system to go back and forth from state a to state B.

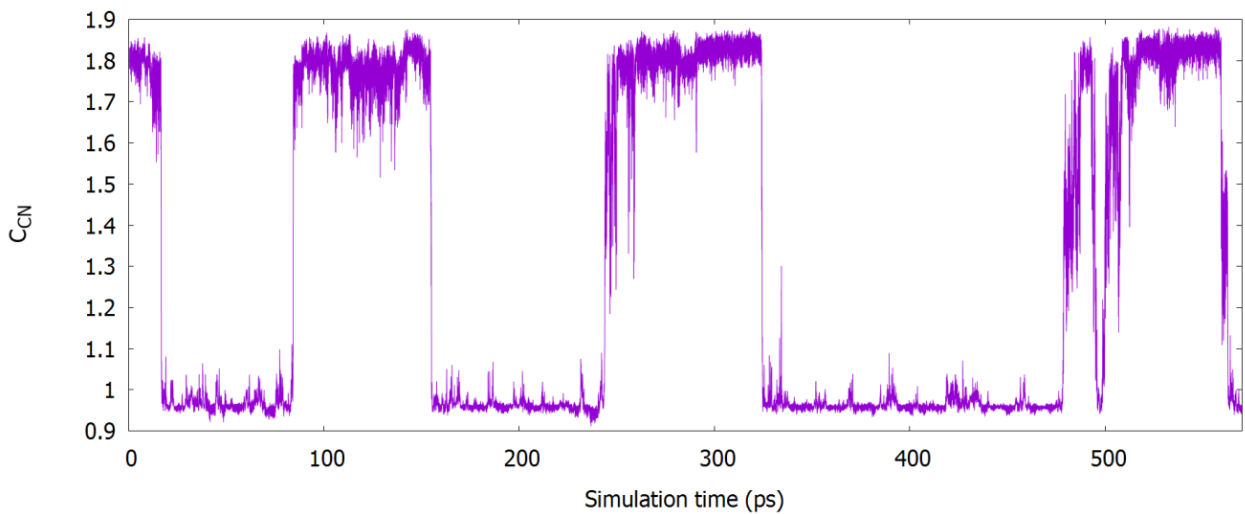

Figure 1. Collective variable $C_{CN}$ as a function of simulation time.

In Figure 2 we show the convergence of the free energy difference between urea (A) and isocyanic acid and ammonia (B), where in the B basin is considered also the dissociated cyanate and ammonium ions, corresponding to the B1 basin.

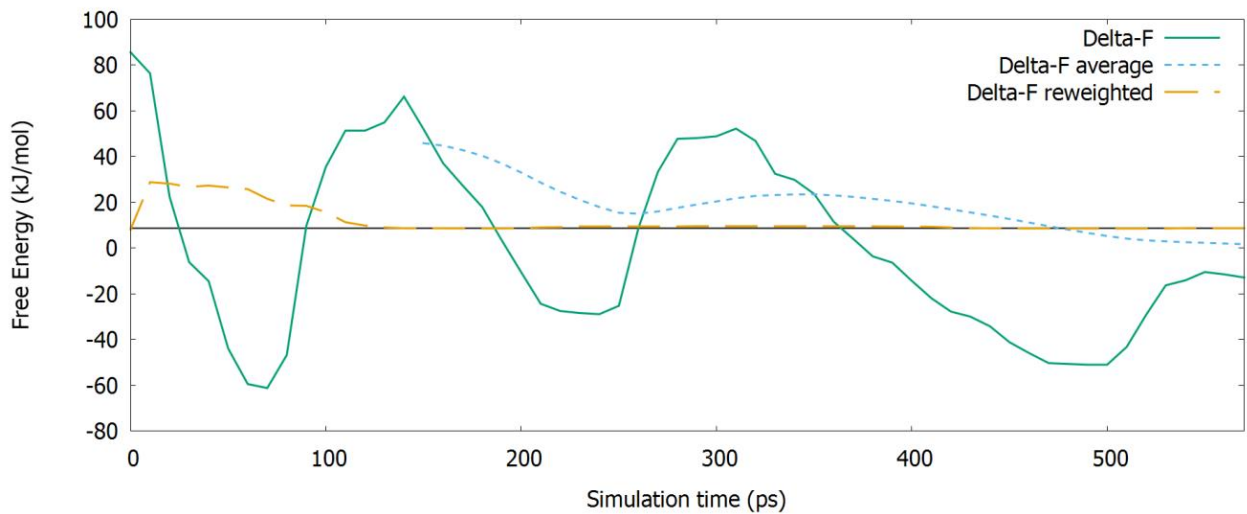

Figure 2. Free energy difference as a function of simulation time.

## 2.2 Static Calculations

Geometry, frequencies, rotational constants and symmetry numbers of reactants, products and transition state calculated at the B3LYP/6-31+G(d,p) level.

### UREA

```
Coordinates XYZ (Angstrom)

  C    -0.000110     0.140106     0.000117
  O    -0.000003     1.364523     0.000011
  N     1.163322    -0.610889    -0.067732
  N    -1.163276    -0.610862     0.067627
  H    -2.004132    -0.069129    -0.073664
  H    -1.172883    -1.533543    -0.344190
  H     2.003945    -0.068602     0.073050
  H     1.173428    -1.533286     0.344744

Rotational constants (GHz): 11.13089    10.28293     5.38534

Rotational Symmetry number: 2

Frequencies (cm-1)

   441.7622              444.9463              494.9305
   548.7583              562.0510              590.8510
   798.8349              952.9709             1053.9188
  1174.0166             1391.0628             1615.2803
  1615.6238             1734.1736             3548.2121
  3552.1084             3665.3965             3666.7142
```

## H2O

Coordinates XYZ (Angstrom)

```
O     0.000000     0.116610     0.000000
H     0.768938    -0.466449     0.000000
H    -0.768938    -0.466430     0.000000
```

Rotational constants (GHz): 830.49369   424.05246   280.71737

Rotational Symmetry number: 2

Frequencies (cm-1)

| | | |
|---|---|---|
| 1613.0238 | 3798.3814 | 3896.9616 |

## HNCO

Coordinates XYZ (Angstrom)

```
C     0.000000     0.051199     0.000000
O    -0.625027     1.045888     0.000000
N     0.501678    -1.058613     0.000000
H     1.488469    -1.264008     0.000000
```

Rotational constants (GHz): 895.43188   10.94230   10.81020
Rotational Symmetry number: 1

Frequencies (cm-1)

| | | |
|---|---|---|
| 561.9900 | 683.2323 | 776.0602 |
| 1314.6840 | 2220.8409 | 3665.8666 |

## NH3

Coordinates XYZ (Angstrom)

```
N     0.000000    -0.000053    -0.107927
H     0.822912    -0.474763     0.251877
H    -0.822830    -0.474903     0.251878
H    -0.000080     0.950040     0.251736
```

Rotational constants (GHz): 299.76309   299.63341   185.19301

Rotational Symmetry number: 3

Frequencies (cm-1)

| | | |
|---|---|---|
| 1008.2102 | 1650.6878 | 1650.8028 |
| 3452.5341 | 3584.4870 | 3585.3969 |

## TRANSITION STATE

Coordinates XYZ (Angstrom)

```
C    -0.838521   -0.148968    0.003550
N    -0.005119    1.172575   -0.009989
O    -2.052346   -0.021702    0.029883
N     0.002252   -1.162289   -0.018033
H    -0.319729    1.749572    0.770257
H    -0.230612    1.677340   -0.867973
H    -0.448237   -2.070143   -0.031776
H     1.238901   -0.823802   -0.028637
O     2.165409   -0.034026   -0.073521
H     1.107477    0.871062    0.035296
H     2.798897   -0.136408    0.646792
```

Rotational constants (GHz):  9.71639    3.09068    2.38611

Optical symmetry number: 2

Frequencies (cm-1)

| | | |
|---|---|---|
| -1327.2448 | 126.2142 | 376.9010 |
| 393.0904 | 469.7945 | 534.7944 |
| 568.9234 | 625.9217 | 654.8309 |
| 697.7538 | 733.8250 | 773.2523 |
| 1094.0858 | 1117.6744 | 1126.7591 |
| 1334.4421 | 1361.2987 | 1397.0239 |
| 1557.5976 | 1632.8666 | 1732.2832 |
| 1777.7268 | 1890.1359 | 3454.1671 |
| 3526.0128 | 3562.3601 | 3845.6232 |